\documentstyle[12pt]{article}
\begin{document}
\vspace{2cm}
\title{Diagonalization of 2-D inhomogeneous model related to
the Hubbard model }
\author{ Ruihong Yue\thanks{email: yue@phys.ocha.ac.jp }  
      ~ and Tetsuo Deguchi \\[.3cm]
      Department of Physics  \\
      Faculty of Science\\
      Ochanomizu University \\
      112 Tokyo, Japan}

\date{}
\maketitle
\begin{abstract}
We found the eigenvalues of the transfer matrix for the 
2-D inhomogeneous statistical model with twisted boundary condition
by using the analytic Bethe Ansatz method. In the uniform case, the 
derived hamiltonian generalizes the 1-D Hubbard model with the 
twisted boundary. We 
also give the energy spectra for the derived hamiltonian. 
\end {abstract}
\vspace{.5cm}\noindent
\raisebox{14cm}[][]{\hspace{2cm} \bf  OCHA-PP-88\\}
\newpage
\bigskip
\section{Introduction}

The Hubbard model is an important  model in condensed matter 
physics. Lieb and Wu \cite{LW}  diagonalized the 1-D Hubbard model 
in terms of the coordinate Bethe Ansatz. The existence of 
the Bethe Ansatz equations implies the integrability of the model. 
In order to prove it from the QISM, Shastry \cite{Sha1,Sha2} proposed 
a coupled 6-vertex model and constructed the transfer matrix 
related to the 1-D Hubbard model through the $R$ matrix and the 
$L$ operator. Olmedilla, Wadati and Akutsu \cite{WOA,OWA,OW} obtained the 
supersymmetric $R$ matrix and $L$ operator by applying the 
Jordan-Wigner transformation to the $L$ operator in \cite{Sha1,Sha2}. 
In \cite{Sha3}, the eigenvalue of the transfer matrix for 
the 1-D Hubbard model was conjectured. Using the coordinate Bethe 
Ansatz method, Bariev \cite{Bar} derived the eigenvalue of the 
diagonal-to-diagonal transfer matrix of the coupled 6-vertex model. 
Recently, we \cite{YD} found the eigenvalues of the transfer matrices 
of the 1-D Hubbard model and the coupled 6-vertex model with twisted 
boundary condition by using the Analytic Bethe Ansatz method.
The eigenvalue of the transfer matrix of the 1-D Hubbard model was also
studied  by Ramos and Martins \cite{RM} from the viewpoint of the 
Algebraic Bethe Ansatz method.

The $R$ matrix of the 1-D Hubbard model has the quite different property.
The   $R(\mu_1,\mu_2)$ matrix of the XXZ model depends on the 
difference of the two spectral parameters $\mu_1$ and $\mu_2$, 
where $\mu_i$ is defined for the $i$-th auxiliary space; If 
$\mu_1-\mu_2 = \mu'_1-\mu'_2$, then $R(\mu_1,\mu_2)=R(\mu'_1,\mu'_2)$.  
For the 1-D Hubbard model, however, this is not the case. Furthermore, the 
$R$ matrix and the $L$ operator of the 1-D Hubbard model are different;
the $R$ matrix depends on two spectral parameters, while the $L$ depends 
on one.  As printed out in \cite{Sha3}, the proof of the Yang-Baxter
equation $RRR=RRR$ is independent from that of $RLL=LLR$.
Shiroishi and Wadati \cite{SW} proved the Yang-Baxter equation $RRR=RRR$
and obtained a new hamiltonian with periodic 
boundary condition. In this paper, we discuss the diagonalization of the 
 inhomogeneous 2-D model with twisted boundary. By applying
the Analytic Bethe Ansatz approach, we find the eigenvalue of the 
transfer matrix and the Bethe Ansatz equations. Under the uniform limit 
(homogeneous case), the 
logarithmic derivative of the eigenvalue of the transfer matrix for the 
2-D model gives the energy spectra of the derived hamiltonian 
with twisted boundary. If the twisted angles are zero,
the hamiltonian is reduced to that  given in \cite{SW}. 

\section{Model}

The $R$ matrix related to the Hubbard model is given in \cite{Sha1,Sha2}
\begin{equation}
\begin{array}{rcl}
R_{12}(\mu_1,\mu_2)&=&\displaystyle
\frac{8\sinh(h_1+h_2)}{U\sin2(\mu_1+\mu_2)\cos(\mu_1-\mu_2)}\\[3mm]
& &\displaystyle\times\left\{\cos(\mu_1+\mu_2)\cosh(h_1-h_2)
L^{\sigma}_{12}(\mu_1-\mu_2)L^{\tau}_{12}(\mu_1-\mu_2) \right.\\[3mm]
& &\displaystyle \left.+\cos(\mu_1-\mu_2)\sinh(h_1-h_2)
L^{\sigma}_{12}(\mu_1+\mu_2)\sigma^z_2
L^{\tau}_{12}(\mu_1+\mu_2)\tau^z_2 \right\},
\end{array}
\end{equation}
where $\sigma$ and $\tau$ are Pauli matrices. The $\mu_j$ are spectral 
parameters and $h_j$ describing the interaction strength. They are 
controlled by $\sinh2h_j=(U\sin\mu_j)/4$. The $L^a$ for $a=\sigma,\tau$ 
are defined by 
\begin{eqnarray}
L^{\sigma}_{12}(\mu)&=&w_4(\mu)+w_3(\mu)\sigma_1^z\sigma_2^z
                       +\sigma_1^+\sigma_2^-+\sigma_1^-\sigma_2^+,\nonumber\\
L^{\tau}_{12}(\mu)&=&w_4(\mu)+w_3(\mu)\tau_1^z\tau_2^z
                       +\tau_1^+\tau_2^-+\tau_1^-\tau_2^+,
\end{eqnarray}
where
\begin{eqnarray}
w_4(\mu)+w_3(\mu)&=&\cos(\mu),\nonumber \\
w_4(\mu)-w_3(\mu)&=&\sin(\mu).
\end{eqnarray}
This $R$ matrix satisfies the Yang-Baxter equation \cite{SW}
\begin{equation}
R_{21}(\mu_1,\mu_2)R_{31}(\mu_1,\mu_3)R_{32}(\mu_2,\mu_3)
=R_{31}(\mu_1,\mu_3)R_{32}(\mu_2,\mu_3)R_{21}(\mu_1,\mu_2).
\end{equation}

The Yang-Baxter equation (4) implies the existence of a general 
2-D inhomogeneous model with the transfer matrix
$$t(\mu,\{\mu_j\})=tr_gT(\mu,\{\mu_j\})=tr_gR_{Lg}(\mu,\mu_L)\cdots
                    R_{1g}(\mu,\mu_1).
$$
The logarithmic derivative of $t(\mu,\{\mu_j\})$ at $\mu=\mu_j=\mu_0$
gives a 1-D quantum system with periodic boundary condition \cite{SW}.
Here, we consider the twisted transfer matrix
\begin{eqnarray}
t(\mu,\{\mu_j\})&=&T_{11}(\mu,\{\mu_j\})e^{-i\phi}
                  +T_{22}(\mu,\{\mu_j\})e^{-i\psi}\nonumber \\[3mm]
                & &+T_{33}(\mu,\{\mu_j\})e^{i\psi}+
                   T_{44}(\mu,\{\mu_j\})e^{i\phi}
\end{eqnarray}
where $\phi$ and $\psi$ are free parameters (twisted angles). It is easy 
to prove that the twisted transfer matrix is the generating function of 
the infinite  number of the conserved quantities.

With the initial condition of $R$ matrix 
\begin{equation}
R_{mg}(\mu_0,\mu_m=\mu_0)=P_{mg}=P^{\sigma}_{mg}P^{\tau}_{mg},
\end{equation}
$\mu_0$ being arbitrary parameter, the hamiltonian related to the twisted
transfer matrix (5) is given by the logarithmic derivative
\begin{equation}
\begin{array}{rcl}
H&=&\displaystyle\sum_{m=1}^{L-1}(\sigma^+_{m+1}\sigma^-_{m}+
    \sigma^-_{m+1}\sigma^+_{m})+
    \sum_{m=1}^{L-1}(\tau^+_{m+1}\tau^-_{m}+
    \tau^-_{m+1}\tau^+_{m}) \\[5mm]
& &\displaystyle +e^{i(\phi+\psi)}\sigma^+_{L}\sigma^-_{1}+
     e^{-i(\phi+\psi)}\sigma^-_{L}\sigma^+_{1}+
    e^{i(\phi-\psi)}\tau^+_{L}\tau^-_{1}+
    e^{i(\psi-\phi)}\tau^-_{L}\tau^+_{1} \\[5mm]
& &\displaystyle +\frac{U}{4\cosh(2h_0)}\sum_{m=1}^{L-1}
   \begin{array}[t]{l}\left\{\cos^2(\mu_0)\sigma^z_m 
          -\sin^2(\mu_0)\sigma_{m+1}^z\right. \\[3mm]
          \;\; \left.  +\sin(2\mu_0)(\sigma^+_{m+1}\sigma^-_{m}
            -\sigma^-_{m+1}\sigma^+_{m})
        \right\}\end{array} \\[5mm]
 & &\times \left\{\cos^2(\mu_0)\tau^z_m-\sin^2(\mu_0)\tau_{m+1}^z
   +\sin(2\mu_0)(\tau^+_{m+1}\tau^-_{m}
   -\tau^-_{m+1}\tau^+_{m})\right\}\\[5mm]
& &\displaystyle +\frac{U}{4\cosh(2h_0)}
   \begin{array}[t]{l}\left\{
          \cos^2(\mu_0)\sigma^z_L-\sin^2(\mu_0)\sigma_{1}^z \right.  \\[3mm] 
          \;\;\left.+\sin(2\mu_0) (e^{i(\phi+\psi)}\sigma^+_{L}\sigma^-_{1}
          -e^{-i(\phi+\psi)}\sigma^-_{L}\sigma^+_{1})
          \right\}\end{array}\\[5mm]
 & &\times \left\{\cos^2(\mu_0)\tau^z_L-\sin^2(\mu_0)\tau_{1}^z
   +\sin(2\mu_0)(e^{i(\phi-\psi)}\tau^+_{L}\tau^-_{1}
   -e^{i(\psi-\phi)}\tau^-_{L}\tau^+_{1})\right\}.
\end{array}
\end{equation}
This hamiltonian has four free parameters, the $U$ and $\mu_0$ denote the 
interaction strength. The $\phi$ and $\psi$ describe the twisted 
boundary. This hamiltonian will reduce into the 1-D coupled XY model under $\mu_0=0$.
Generally, the hamiltonian (7) contains the interaction between the
charge sector and spin sector. This is due to that the $R$ matrix
does not depends on the difference of two spectral parameters. 
Notice that  this hamiltonian will recover the one in \cite {SW} 
at $\phi=\psi=0$. 

\section{Diagonalization}

In \cite{YD}, we obtained the eigenvalues of the transfer matrices 
related to the 1-D Hubbard model and the coupled twisted XY model by 
using the Analytic Bethe Ansatz method. Here we want to find the eigenvalue
of the transfer matrix (5) by taking use of the same idea. Let us define
the total reference state to be the state with all spins down (spin 
$\sigma$ and spin $\tau$). The matrix $T(\mu,\{\mu_j\})$ on the reference
state takes the form
\begin{equation}
T(\mu,\{\mu_j\})|vac>=\left[\begin{array}{cccc}
A_1(\mu)& 0&0&0\\
 \mbox{$*$} &A_2(\mu)&0&0\\
 \mbox{$*$} &0&A_2(\mu)&0\\
 \mbox{$*$} & * & * &A_4(\mu)\end{array}\right]|vac>,
\end{equation}
where $* $ stands for the no-vanishing terms and 
\begin{eqnarray}
         A_1(\mu)&=&\displaystyle\prod_{j=1}^L
                    \rho_8(\mu,\mu_j),\nonumber \\ 
         A_2(\mu)&=&\displaystyle\prod_{j=1}^L
                    \rho_9(\mu,\mu_j),\nonumber \\
         A_4(\mu)&=&\displaystyle\prod_{j=1}^L
                    \rho_1(\mu,\mu_j),\nonumber \\
\rho_8(\mu,\mu_j)&=&\displaystyle\frac{e^{h-h_j}\cos(\mu)\cos(\mu_j)
                    -e^{h_j-h}\sin(\mu)\sin(\mu_j)}
                    {\cos^2(\mu)-\sin^2(\mu_j)}\nonumber\\   
                 & &-e^{h-h_j}\cos(\mu)\cos(\mu_j)
                    -e^{h_j-h}\sin(\mu)\sin(\mu_j),\nonumber \\
\rho_9(\mu,\mu_j)&=&\displaystyle \sin(\mu-\mu_j)\cosh(h-h_j)
                    -\sin(\mu+\mu_j)\sinh(h-h_j),\nonumber \\
\rho_1(\mu,\mu_j)&=&\displaystyle \cos(\mu-\mu_j)\cosh(h-h_j)
                    +\cos(\mu+\mu_j)\sinh(h-h_j).
\end{eqnarray}
Using the explicit expression of the $R$ matrix, one can easily find the 
eigenvalues of the states with $N$ $\tau$-spin (or $\sigma$-spin) 
flipping from the reference state.
For the 1-D Hubbard model, $N$ corresponds to the number of electrons.
 After a long but direct calculation, we arrive at 
\begin{equation}
\begin{array}{rcl}
\Lambda_N(\mu)&=&\displaystyle A_4(\mu)e^{i\phi}\prod_{j=1}^N
                 \frac{\rho_1(\nu_j,\mu)}{\rho_9(\nu_j,\mu)}
                 +A_2(\mu)e^{i\psi}\prod_{j=1}^N
                 \frac{\rho_4(\mu,\nu_j)}{\rho_9(\mu,\nu_j)}\\[3mm]
              & &\displaystyle +A_2(\mu)e^{-i\psi}\prod_{j=1}^N
                 \frac{\rho_{10}(\mu,\nu_j)}
                  {\rho_1(\mu,\nu_j)-\rho_3(\mu,\nu_j)}\\[3mm]
              & &\displaystyle +A_1(\mu)e^{-i\phi}\prod_{j=1}^N
                 \frac{\rho_{10}(\mu,\nu_j)}
                  {\rho_3(\mu,\nu_j)-\rho_1(\mu,\nu_j)},
\end{array}
\end{equation}
where
\begin{equation}
\begin{array}{rcl}
\rho_3(\mu,\nu_j)&=&\displaystyle \frac{e^{h-\hat{h}_j}\cos(\mu)\cos(\nu_j) 
                    -e^{\hat{h}_j-h}\sin(\mu)\sin(\nu_j)}
                     {\cos^2(\mu)-\sin^2(\nu_j)}, \\[3mm]
\rho_4(\mu,\nu_j)&=&\displaystyle e^{\hat{h}_j-h}\cos(\mu)\cos(\nu_j) 
                    +e^{h-\hat{h}_j}\sin(\mu)\sin(\nu_j), \\[3mm]
\rho_6(\mu,\nu_j)&=&\displaystyle \frac{e^{-2\hat{h}_j}\cos(\mu)\sin(\mu) 
                    -e^{-2h}\cos(\nu_j)\cos(\nu_j)}
                     {\cos^2(\mu)-\sin^2(\nu_j)}, \\[3mm]
\rho_3(\mu,\nu_j)&=&\displaystyle e^{h-\hat{h}_j}\sin(\mu)\cos(\nu_j) 
                    -e^{\hat{h}_j-h}\cos(\mu)\sin(\nu_j) .
\end{array}
\end{equation}
where the notation $\sinh(2\hat{h}_j)=U\sin(2\nu_j)/4$ has been used.

In accordance with the hypothesis of the Analytic Bethe Ansatz approach 
\cite{YD}, one now seeks for a more general  form
\begin{equation}
\begin{array}{rcl}
\Lambda_N(\mu)&=&\displaystyle A_4(\mu)e^{i\phi}\prod_{j=1}^N
                 \frac{\rho_1(\nu_j,\mu)}{\rho_9(\nu_j,\mu)}\\
              & &\displaystyle +A_2(\mu)e^{i\psi}\prod_{j=1}^N
                 \frac{\rho_4(\mu,\nu_j)}{\rho_9(\mu,\nu_j)}
                 \prod_{m=1}^Mg_3(\mu,\lambda_m)\\[3mm]
              & &\displaystyle +A_2(\mu)e^{-i\psi}\prod_{j=1}^N
                 \frac{\rho_{10}(\mu,\nu_j)}
                  {\rho_1(\mu,\nu_j)-\rho_3(\mu,\nu_j)}
                  \prod_{m=1}^M g_2(\mu,\lambda_m) \\[3mm]
              & &\displaystyle +A_1(\mu)e^{-i\phi}\prod_{j=1}^N
                 \frac{\rho_{10}(\mu,\nu_j)}
                  {\rho_3(\mu,\nu_j)-\rho_1(\mu,\nu_j)},
\end{array}
\end{equation}
where $g_2(\mu,\lambda_m)$ and $g_3(\mu,\lambda_m)$ are some undetermined 
functions. Here $N$ is the total number of $\sigma$ up-spins and $\tau$ 
up-spins, $M$ the number of $\tau$ up-spins. Because the $R$ matrix
has no crossing symmetry, these functions can not be fixed by using the 
standard Analytic Bethe Ansatz method. Let us consider some properties of 
these functions.  The eigenvalue $\Lambda(\mu)$ as the analytic function 
of $\mu$ should have vanishing residues at the poles $\mu=\nu_j$ and 
$\mu=\tilde{\nu}_j$, which sets up the relation
\begin{equation}
\prod_{m=1}^Mg_3(\nu_j,\lambda_m)=
\prod_{m=1}^Mg_2(\tilde{\nu}_j,\lambda_m)
\end{equation}
where $e^{-2\tilde{h}_j}\cot(\tilde{\nu}_j)=e^{2\hat{h}}\cot(\nu_j)$. 
Second, from the special case of $\Lambda$ at $N=2$, $M=1$, we know that 
the functions $g_2$ and $g_3$ are the rational and have some poles. In 
terms of the news variables $k$ defined by $e^{ik}=-e^{-2h}\cot(\mu)$, $g_2$
and $g_3$ can be written as 
\begin{equation}
g_2(\mu,\lambda)=\frac{P_2(\mu,\lambda)}{i\sin(k)-\lambda+U/4}\;,\;
g_3(\mu,\lambda)=\frac{P_3(\mu,\lambda)}{i\sin(k)-\lambda+U/4}
\end{equation}
where $P_2$ and $P_3$ are analytic function of $\mu$. The equation (13), 
together with the asymptotic behavior of $\Lambda(\mu)$, fixes completely 
$P_2$ and $P_3$. Thus, we obtain the final result
\begin{equation}
\begin{array}{rcl}
\Lambda_N(\mu)&=&\displaystyle A_4(\mu)e^{i\phi}\prod_{j=1}^N
                 \frac{\rho_1(\nu_j,\mu)}{\rho_9(\nu_j,\mu)} \\
              & &\displaystyle +A_2(\mu)e^{i\psi}\prod_{j=1}^N
                 \frac{\rho_4(\mu,\nu_j)}{\rho_9(\mu,\nu_j)}
                 \prod_{m=1}^M-\frac{i\sin(k)-\lambda_m-U/4}
                       {i\sin(k)-\lambda_m+U/4}\\[3mm]
              & &\displaystyle +A_2(\mu)e^{-i\psi}\prod_{j=1}^N
                 \frac{\rho_{10}(\mu,\nu_j)}
                  {\rho_1(\mu,\nu_j)-\rho_3(\mu,\nu_j)}
                  \prod_{m=1}^M -\frac{i\sin(k)-\lambda_m+3U/4}
                       {i\sin(k)-\lambda_m+U/4} \\[3mm]
              & &\displaystyle +A_1(\mu)e^{-i\phi}\prod_{j=1}^N
                 \frac{\rho_{10}(\mu,\nu_j)}
                  {\rho_3(\mu,\nu_j)-\rho_1(\mu,\nu_j)},
\end{array}
\end{equation}
the Bethe Ansatz equations are 
\begin{equation}
\begin{array}{rcl}
\displaystyle  e^{i(\phi-\psi)}\frac{A_4(\nu_j,\{\mu_l\})}
   {A_2(\nu_j,\{\mu_l\})}
&=&\displaystyle (-1)^{M+N+1+L}
   \prod_{m=1}^M\frac{i\sin(k_j)-\lambda_m-U/4}{i\sin(k_j)-\lambda_m+U/4}
    \\[5mm]
\displaystyle\prod_{j=1}^N
 \frac{i\sin(k_j)-\lambda_n+U/4}{i\sin(k_j)-\lambda_n-U/4}
&=&\displaystyle (-1)^{M+1}e^{i2\psi}
   \prod_{m=1}^M\frac{\lambda_n-\lambda_m-U/2}{\lambda_n-\lambda_m+U/2}
,\end{array}
\end{equation}
where $e^{ik_j}=-e^{-2\hat{h}_j}\cot(\nu_j)$. Equations (15) and 
 (16) are the eigenvalue and the Bethe Ansatz equations
for the 2-D inhomogeneous model. If the solutions of equation (16) 
are given, we can find the eigenvalues of $\Lambda$. Furthermore,
 we can calculate the thermodynamics of the system.

Now, let us return to consider the energy spectra of the hamiltonian (7). 
Taking the logarithmic derivative of 
$\Lambda(\mu)$ at $\mu=\mu_0$ and letting $\mu_j=\mu_0$, we get
\begin{equation}
\begin{array}{rcl}
E&=&\displaystyle \frac{\cos^2(2\mu_0)UL}{4}
    -\frac{NU}{2\cosh(2h_0)} -2\sum^N_{j=1}\cos(k_j)
    +\frac{U}{2}\sum_{j=1}^N\\[5mm]
 & &\displaystyle \left\{\frac{\cosh^{-1}(2h_0)[\sinh(2h_0-2\hat{h}_j)
                         -\cos(2\nu_j)\cosh(2h_0-2\hat{h}_j)]}
         {\cot(2\mu_0)\sin(2\nu_j)+\sinh(2h_0-2\hat{h}_j)
                 -\cos(2\nu_j)\cosh(2h_0-2\hat{h}_j)}\right.\\[5mm]        
 & &\displaystyle - \frac{[\sinh(2h_0-2\hat{h}_j) 
              -\cos(2\nu_j)\cosh(2h_0-2\hat{h}_j)]}
       {[\cot(2\mu_0)\sin(2\nu_j)+\sinh(2h_0-2\hat{h}_j)
                 -\cos(2\nu_j)\cosh(2h_0-2\hat{h}_j)]}\\[5mm]
& & \displaystyle \;\;\;\times 
    \frac{[\cosh(2\hat{h}_j)+\cos(2\nu_j)\sinh(2\hat{h}_j)]}
    {\sin(2\nu_j)}    \\[5mm]
& &\displaystyle \left..-
    \frac{\sinh(2\hat{h}_j)+\cos(2\nu_j)\cosh(2\hat{h}_j)}
         {\cot(2\mu_0)\sin(2\nu_j)+\sinh(2h_0-2\hat{h}_j)
                 -\cos(2\nu_j)\cosh(2h_0-2\hat{h}_j)}\right\}.
 \end{array}
\end{equation}

\section{conclusion}

We have found the eigenvalue and the Bethe Ansatz equations for the  2-D 
inhomogeneous twisted model by making use of the Analytic Bethe Ansatz 
approach. We have also obtained  the energy spectra (17) for the related 1-D 
quantum hamiltonian (7).  In equation (17), the first three terms are similar
to that  of the 1-D Hubbard model. The terms in the curved bracket reflect the 
contribution due to  the interaction between the spin and  charge 
sectors. In the $R(\mu_1,\mu_2)$ matrix, the spectral shift does not keep 
the invariance of $R$, i.e. $R(\mu_1+\delta,\mu_2+\delta)\neq R(\mu_1,\mu_2)$.
Thus the interaction between the two sectors can not removed by using the 
spectral shift. The uniform shift of $\nu_j$'s can not give the uniform shift 
of $k_j$'s due to the nonlinear relation between $\nu_j$'s
and $k_j$'s. This is very different from the case in which the $R(\mu_1,\mu_2)$ 
just depends on the difference of the two spectral parameters.

In fact, the eigenvalues of the 1-D Hubbard model and  the coupled XY model
can be considered as the special case of equation (17). 
The $L$ operator of the 1-D Hubbard model is obtained from the 
$R(\mu_1,\mu_2)$ by setting $\mu_2=0$ in \cite{Sha1,Sha2}. 
This means the uniform parameter $\mu_0$ 
being zero. At this case, the equations (16) and (17) recover the
results given in \cite{YD}.

Based upon the equations (15)-(17), one can calculate the finite-size 
correction and get the conformal scales. 
Notice that the twisted angles $\phi$ and $\psi$ can be 
interpreted as the external vector potentials coupled to the system, 
one can consider the conductivity of the system as done in
1-D Hubbard model and 1-D t-J model \cite{SS,KY}.

{\bf acknowledgement}

 R.Y. was granted by the JSPS foundation and the Monbusho Grand-in-Aid
of Japanese Government.

\end{document}